\documentclass[aps,prc,letterpaper,11pt,twoside,tightenlines,nofootinbib,showpacs,preprint]{revtex4}
\usepackage{graphicx}
\usepackage[sort&compress]{natbib}
\usepackage{subfigure}
\usepackage{amsmath}
\usepackage{amsfonts}
\usepackage{cancel}

\begin{document}

\arraycolsep1.5pt

\newcommand{\Ima}{\textrm{Im}}
\newcommand{\Rea}{\textrm{Re}}
\newcommand{\mev}{\textrm{ MeV}}
\newcommand{\be}{\begin{equation}}
\newcommand{\ee}{\end{equation}}
\newcommand{\ba}{\begin{eqnarray}}
\newcommand{\ea}{\end{eqnarray}}
\newcommand{\gev}{\textrm{ GeV}}
\newcommand{\nn}{{\nonumber}}
\newcommand{\dtres}{d^{\hspace{0.1mm} 3}\hspace{-0.5mm}}

\title{A description of the $f_2(1270)$, $\rho_3(1690)$,   $f_4(2050)$,
 $\rho_5(2350)$ and   $f_6(2510)$ resonances as multi-$\rho(770)$ states 
}

\author{L.~Roca}
\affiliation{Departamento de F\'{\i}sica. Universidad de Murcia. E-30071, Murcia. Spain}

\author{E.~Oset}
\affiliation{
Departamento de F\'{\i}sica Te\'orica and IFIC, Centro Mixto Universidad de Valencia-CSIC,
Institutos de Investigaci\'on de Paterna, Aptdo. 22085, 46071 Valencia,
Spain}

\date{\today}

 \begin{abstract} 
In a previous work regarding the interaction 
of two $\rho(770)$ resonances, the $f_2(1270)$ ($J^{PC}=2^{++}$) resonance was obtained
dynamically as a two-$\rho$ molecule with a very strong
 binding energy, 135~MeV per $\rho$ particle. 
In the present work we use the $\rho\rho$ interaction
in spin 2 and isospin 0 channel
to show that the resonances 
$\rho_3(1690)$ ($3^{--}$),   $f_4(2050)$ ($4^{++}$),
 $\rho_5(2350)$ ($5^{--}$) and   $f_6(2510)$ ($6^{++}$)
are basically molecules of increasing number of $\rho(770)$ particles.
We use the fixed center approximation of the Faddeev equations to write
the multi-body interaction in terms  of the two-body scattering
amplitudes.
We find the masses of the states very close to the experimental values
and we get an increasing value of the binding energy per $\rho$ 
as the number of $\rho$ mesons is increased.

\end{abstract}

\maketitle

\section{Introduction}
\label{Intro}

In the last decade, the chiral unitary approach has shown that many
hadronic resonances can be obtained dynamically from the interaction of
hadrons. This has been done through the implementation of unitarity in
coupled channels using a lowest order chiral  Lagrangian, to the point
that these resonances can be interpreted as meson-meson or meson-baryon
molecules
\cite{Kaiser:1995cy,npa,iam,nsd,Kaiser:1998fi,angels,juanenrique,ollerulf,carmenjuan,hyodo},
and it has shed new light into the issue of the nature of the scalar
mesons, among others.  The interaction of pseudoscalar mesons among
themselves and meson-baryon interaction has given way recently to
the interaction of vector mesons among themselves
\cite{Molina:2008jw,gengvec,nagahiro} and the interaction of vector
mesons with baryons \cite{sourav,angelsvec}, where the interaction is
evaluated within the techniques of the chiral unitary approach starting
from a lowest order hidden gauge symmetry Lagrangian
\cite{hidden1,hidden2,hidden3,hidden4}. In the work \cite{Molina:2008jw}
it was found that the interaction of two $\rho(770)$ mesons in isospin
$I=0$ and spin $S=2$ was strong enough to bind the $\rho\rho$ system
into the $f_2(1270)$ ($J^{PC}=2^{++}$) resonance.  The nature of this
resonance as a   $\rho(770)\rho(770)$ molecule has passed the tests of
radiative decay into $\gamma \gamma$ \cite{junko}, the decay of $J/\Psi$
into $\omega (\phi)$ and $f_2(1270)$ (together with other resonances
generated in \cite{gengvec})  \cite{daizou}, and $J/\Psi$ into $\gamma$
and  $f_2(1270)$ (and the other resonances of \cite{gengvec})
\cite{hanhart}.

The  $f_2(1270)$ obtained in ref.~\cite{Molina:2008jw} as a $\rho\rho$
quasibound state or molecule implies a very large binding energy per
$\rho$ meson, about $135$~MeV. This occurs only for spin S=2, where the
two spins of the $\rho$ are aligned in the same direction. In view of 
this strong $\rho\rho$ interaction, some natural questions  arise: i)
is it possible to obtain bound systems with increasing  number of $\rho$
mesons as building blocks? These systems with many $\rho$'s with their
spins aligned in the same direction would make a condensate, with
features similar to a ferromagnet; ii) If so, is there a limit in the
number of $\rho$'s or, even more interesting, the mass of the
multi-$\rho$ system saturates at some number of $\rho$ mesons?.  In this
latter case then it would be energetically "free" to introduce new
$\rho$ mesons into the system.

  The condensates made out of mesons have been advocated some times, and concretely, the issue of pion condensates was popular for some time 
 \cite{Migdal:1974jn} and kaon condensation has also attracted much attention \cite{Kaplan:1986yq}.
  
Regarding the question i), in the PDG~\cite{Amsler:2008zzb}
there are intriguing mesons with large spin, of the $\rho$ and $f_0$
type, whose quantum numbers match systems made with 3, 4, 5 and 6 $\rho$ mesons with their spins aligned. 
These are the $\rho_3(1690)$ ($3^{--}$),   $f_4(2050)$ ($4^{++}$),
 $\rho_5(2350)$ ($5^{--}$) and   $f_6(2510)$ ($6^{++}$)
resonances. If these resonances were essentially 
multi-$\rho$ meson molecules, they would have a binding energy per
$\rho$ of about 210, 260, 305 
and 355~MeV, respectively. This increasing value as more $\rho$'s are added to the system
connects with question ii).

The main aim of the present work is to address these questions.
Technically, we would have to solve the Faddeev equations
\cite{Faddeev:1960su} for a state of three $\rho$'s to start with.  
Systems of three mesons, concretely $K\bar K\phi$, have been addressed
recently with Faddeev equations  \cite{MartinezTorres:2008gy}. 
States of two mesons and one baryon have also received recent attention
\cite{MartinezTorres:2007sr,MartinezTorres:2008kh,Jido:2008kp,KanadaEn'yo:2008wm}.
In the three $\rho$ mesons state,
the fact that the two $\rho$ meson system with S=2 is so bound, makes it
advisable to use the fixed center approximation to the Faddeev equations
(FCA) \cite{Chand:1962ec,Barrett:1999cw,Deloff:1999gc,Kamalov:2000iy} in
order to obtain the scattering amplitudes of one $\rho$ with the
$f_2(1270)$ state. The FCA requires the knowledge of the wave function
of the bound state of the target. This information can be obtained using
a recent method that connects, in an easy and practical way, the wave
functions with the scattering matrices of the chiral unitary approach
\cite{Gamermann:2009uq}. Proceeding by iterations we build up states
with an extra $\rho$ meson starting from the former state. In this way
the multi-$\rho$ resonances are generated, which show up as prominent
bumps in the different scattering amplitudes.  The iterative method
suggest a way to extrapolate to many $\rho$ states and we develop an
analytical method to evaluate the mass of the $n$-$\rho$ state for very
large $n$ if only the  single-scattering contribution is considered in
the Faddeev equations, which should be viewed only as suggestive of what
might happen in the limit of a large number of $\rho$ mesons.

\section{unitarized $\rho\rho$ interaction}
\label{sec:rhorho}

The most important ingredient in the calculation of the multi-$\rho$
scattering is the two-$\rho$ interaction. In this section we briefly
summarize
the model of ref.~\cite{Molina:2008jw} to obtain the unitary
$\rho\rho$ scattering amplitude. (We refer to \cite{Molina:2008jw} for
further details).

The $\rho\rho$ potential is obtained from the hidden gauge
symmetry Lagrangian \cite{hidden1,hidden2,hidden3,hidden4} 
for vector mesons,
which, up to three and four vector fields, reads:

\begin{equation}
{\cal L}^{(4V)}=\frac{g^2}{2}\langle V_\mu V_\nu V^\mu V^\nu-V_\nu V_\mu
V^\mu V^\nu\rangle\ ,
\label{lcont}
\end{equation}
\begin{equation}
{\cal L}^{(3V)}=ig\langle (V^\mu \partial_\nu V_\mu -\partial_\nu V_\mu V^\mu)  V^\nu\rangle
\label{l3V}\ ,
\end{equation}
where  $V_\nu$ is the $SU(3)$ matrix containing the vector-meson fields
and the coupling constant $g$ is $g=M_V/2f$ with $f=93$\mev, the pion
decay constant.
The Lagrangian of Eq.~(\ref{lcont}) gives rise to a four vector meson contact
term and that of Eq.~(\ref{l3V}) to a four vector meson interaction through the
exchange of an intermediate vector meson in the $t$ and $u$ channels (the s-channel gives rise to a p-wave that we do not consider, only the important s-wave part is studied).

From these Lagrangians, a potential $V$ can be obtained
to which the contact and $\rho$-exchange terms
contribute. For the present work only the spin $S=2$ and isospin $I=0$,
$I=2$, are necessary:

\ba
V^{(I=0,S=2)}(s)&=& -4g^2-8g^2\left(\frac{3s}{4 m_\rho^2}-1\right)
\sim-20g^2\nn\\
V^{(I=2,S=2)}(s)&=&  2g^2+4g^2\left(\frac{3s}{4 m_\rho^2}-1\right)
\sim 10g^2
\ea
where the last terms are the approximate values at threshold in order to
give an idea of the weight and sign of the interaction. The $\rho\rho$
$S=2$, $I=0$ is strongly attractive. This is the most important reason
to obtain a bound $\rho\rho$ state with these quantum
numbers as we explain below.

Further contributions to the previous potential where
considered in ref.~\cite{Molina:2008jw}, out of which only the 
box diagram, which accounts for the two-pion decay mode,
was relevant, and only for the imaginary part of the potential.
 Explicit
expressions can be found in ref.~\cite{Molina:2008jw}.

With this potential the total $\rho\rho$ scattering amplitude can be
obtained. In order to extend the range of
applicability of the interaction to the resonance region,
 the implementation of exact unitarity
is mandatory. In this case, we use the Bethe-Salpeter equation where the
kernel is the potential $V$ described above:
\be
T=\frac{V}{1-VG},
\label{eq:BS}
\ee 
for each spin-isospin channel. In Eq.~\ref{eq:BS}, $G$ is the unitary 
bubble
 or the $\rho\rho$ loop function \cite{npa,iam}

\begin{equation}
G(s)=i\int \frac{d^4
p}{(2\pi)^4}\frac{1}{p^2-m_\rho^2+i\epsilon}\frac{1}{(Q-p)^2-m_\rho^2+i\epsilon}\ ,
\label{loop}
\ee
with $Q=(\sqrt{s},\vec 0)$.
This loop function can be regularized by means of dimensional
regularization or using a three-momentum cutoff, $p_\textrm{max}\equiv
\Lambda$:

\begin{equation}
G(s,m_1,m_2)=\int_0^{\Lambda} \frac{p^2 dp}{(2\pi)^2} 
\frac{\omega_1+\omega_2}{\omega_1\omega_2 
[{(Q^0)}^2-(\omega_1+\omega_2)^2+i\epsilon]   }
\label{eq:Gcutoff}
\end{equation}
 where $\omega_i=(\vec{p}\,^2_i+m_i^2)^{1/2}$.
 
 In order to consider the width of the $\rho$ particles inside
the loop, a convolution with the two $\rho$ meson spectral functions is done 
to Eq.~(\ref{eq:Gcutoff}):

\ba
G(s)= \frac{1}{{\cal N}^2}\int_{(m_\rho-2\Gamma_{ON})^2}^{(m_\rho+2\Gamma_{ON})^2}ds_1
\int_{(m_\rho-2\Gamma_{ON})^2}^{(m_\rho+2\Gamma_{ON})^2}ds_2
\,G(s,\sqrt{s_1},\sqrt{s_2})\, S_\rho(s_1)\, S_\rho(s_2)
\label{eq:Gs}
\ea
where $S_\rho(s_i)$ is the $\rho$ meson spectral function
\be
S_\rho(s_i)=-\frac{1}{\pi}Im\left[\frac{1}{s_i-m_\rho^2
+i m_\rho \Gamma_\rho(\sqrt{s_i})}\right],
\ee
with $\Gamma_\rho(\sqrt{s})$ the $\rho$-meson energy dependent width
\be
\Gamma_\rho(\sqrt{s})=\Gamma_{ON}\left(
\frac{s-4 m_\pi^2}{m_\rho^2-4 m_\pi^2}\right)^{3/2}
\ee
and $\Gamma_{ON}$ the on-shell $\rho$ meson width.
In Eq.~(\ref{eq:Gs}) ${\cal N}$ is a normalization factor given by
\be
{\cal N}=\int_{(m_\rho-2\Gamma_{ON})^2}^{(m_\rho+2\Gamma_{ON})^2}ds
\,S_\rho(s)
\ee

The cutoff $\Lambda$ is the only free parameter in the
 whole model and is
chosen such as to produce the peak of $|T|^2$ at the experimental mass of
the $f_2(1270)$. This implies $\Lambda\simeq 875$\mev, which is of a
natural size \cite{ollerulf}, about 1\gev.

 \begin{figure}[!t]
\begin{center}
\includegraphics[width=0.7\textwidth]{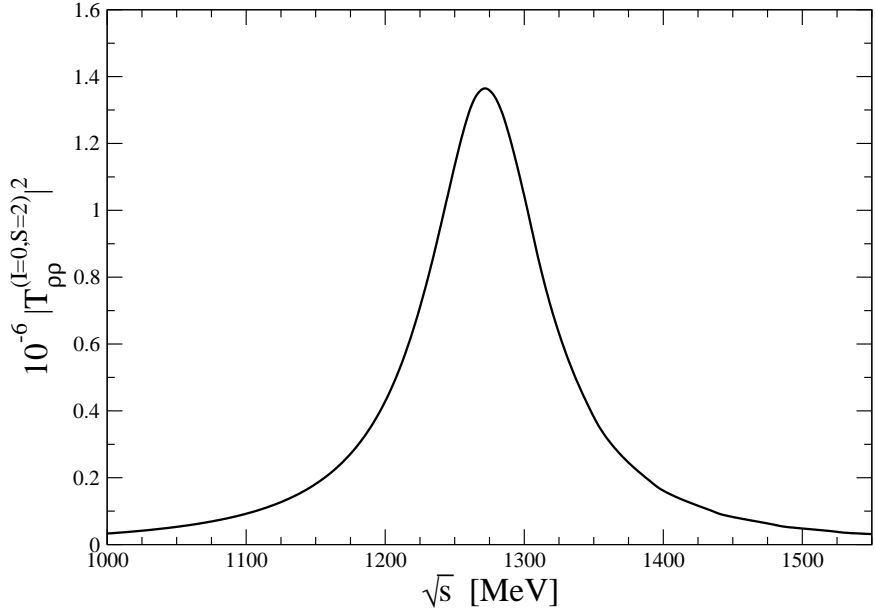}
\caption{Modulus squared of the $\rho\rho$ scattering amplitude 
with total spin
 $S=2$ and
isospin $I=0$}
\label{fig:T2f2}
\end{center}
\end{figure}

In fig.~\ref{fig:T2f2}, the modulus  squared of the $S=2$, $I=0$
scattering amplitude, $T^{(I=0,S=2)}$, is plotted. The
resonance structure of the $f_2(1270)$ resonance is clearly visible.

\section{Multi-body interaction}
\label{sec:Faddeev}

We are going to use the fixed center approximation of the Faddeev
equations in order to obtain the interaction of a number of $\rho$
mesons larger than two.

We will illustrate the process for the interaction of three mesons and 
will give the expression obtained analogously for other number of
mesons.
For the three $\rho$ system, we will consider that
two of the $\rho$ mesons
are clusterized forming an $f_2(1270)$ resonance, given the strong
binding of the $f_2(1270)$ system. This allows us to use the FCA to the Faddeev equations.

The FCA to Faddeev equations is depicted diagrammatically in fig.
\ref{fig:Faddeev}. The external particle, the $\rho$ in this case,
interacts successively with the other two $\rho$ mesons which form the
$\rho \rho$ cluster.  The FCA equations are written in terms of two
partition functions $T_1$, $T_2$, which sum up to the total scattering
matrix, $T$, and read
\ba
T_1&=& t_1+t_1 G_0 T_2 \nn \\
T_2&=& t_2+t_2 G_0 T_1 \nn \\
T&=&T_1+T_2
\ea
where $T$ is the total scattering amplitude we are looking for, $T_i$
accounts for all the diagrams starting with the interaction
of the external particle
 with particle $i$ of the compound
system and $t_i$ represent the $\rho\rho$ unitarized scattering 
amplitude of a $\rho^+$ with any of the other $\rho$ in the $I=0$ $\rho\rho$ system.
The schematic representation is
depicted in fig.~\ref{fig:Faddeev}.

 \begin{figure}[!t]
\begin{center}
\includegraphics[width=0.8\textwidth]{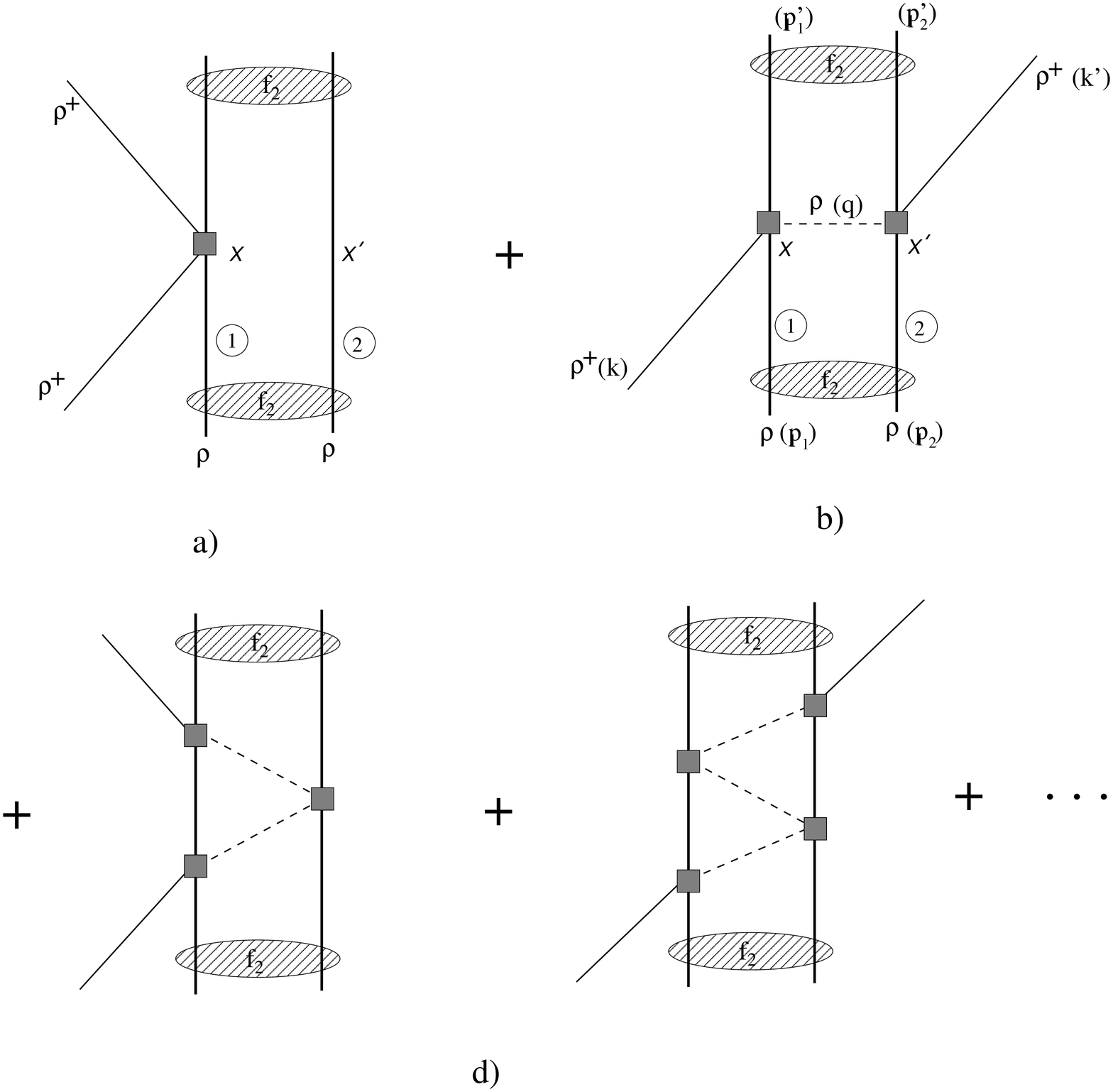}
\caption{Diagrammatic representation of the fixed center approximation
to the Faddeev equations. Diagrams $a)$ and $b)$ represent
the single and double scattering contributions  respectively.}
\label{fig:Faddeev}
\end{center}
\end{figure}

Fig.~\ref{fig:Faddeev}a) represents the single-scattering contribution
and fig.~\ref{fig:Faddeev}b)
the double-scattering. 
The contributions of fig.~\ref{fig:Faddeev}a and b
are the two first
contributions of the Faddeev equations.

In the present case, since both 1 and 2 are $\rho$ mesons we have
$T_1=T_2$ an thus the system of equations is just reduced
 to a single equation

 \ba
T_1&=& t_1+t_1 G_0 T_1 \nn \\
T&=&2T_1
\label{eq:Faddeevred}
\ea

\subsection{Single-scattering contribution}

The amplitude corresponding to the single-scattering contribution of 
fig.~\ref{fig:Faddeev}a comes just from the
$t_1$ term of Eq.~(\ref{eq:Faddeevred}), $T=2 t_1$.

In order to write this expression  
in terms of the $I=0$ and $I=2$ 
unitarized amplitudes ($t_{\rho\rho}^{(I=0)}$, $t_{\rho\rho}^{(I=2)}$)
 of 
Eq.~(\ref{eq:BS}),
let us consider a cluster of two $\rho$ mesons in isospin $I=0$, 
the
constituents of which 
we call mesons 1 and 2. The other $\rho$ meson will be
meson number 3.
The two $\rho$ mesons forming the $f_2$ are in an $I=0$ state 
\be
|\rho\rho\rangle_{I=0}=-\frac{1}{\sqrt{3}}
|\rho^+\rho^- + \rho^-\rho^+ + \rho^0\rho^0\rangle
=\frac{1}{\sqrt{3}}\Big( |(1,-1)\rangle 
+ |(-1,1)\rangle - |(0,0)\rangle\Big)
\ee
where the kets in the last member indicate the $I_z$ components of the 
1 and 2 particles, $|(I_z^{(1)},I_z^{(2)})\rangle$.
We take the $\rho$ meson number 3 in the state $|(I_z^{(3)})\rangle$
\be
|\rho^+\rangle=-|( +1)\rangle.
\ee

The scattering potential in terms of the two body potentials
$V_{31}$, $V_{32}$ is:

\ba
T&= &\left( -\langle (+1)| \otimes \frac{1}{\sqrt{3}} \left(\langle (+1,-1)+
 (-1,+1) -  (0,0)|\right)\right) (V_{31} + V_{32})\nn \\
& & \left(-|(+1)\rangle \otimes \frac{1}{\sqrt{3}} \left(|(+1,-1) +
 (-1,+1) - (0,0)\rangle\right) \right)\nn\\
 &=&\frac{1}{3}
 \bigg\langle ( (2 +\!\!2),-1) + 
 (\frac{1}{\sqrt{6}}(2 0)
         +\frac{1}{\sqrt{2}}(1 0)+\frac{1}{\sqrt{3}}(0 0),1)
  -(\frac{1}{\sqrt{2}}(2 +\!\!1)
         +\frac{1}{\sqrt{2}}(1 +\!\!1),0)
	 \bigg| V_{31}\nn \\
& \bigg|&( (2 +\!\!2),-1) + 
 (\frac{1}{\sqrt{6}}(2 0)
         +\frac{1}{\sqrt{2}}(1 0)+\frac{1}{\sqrt{3}}(0 0),1)
  -(\frac{1}{\sqrt{2}}(2 +\!\!1)
         +\frac{1}{\sqrt{2}}(1 +\!\!1),0)\bigg\rangle
	 \nn\\
&+& \frac{1}{3}
 \bigg\langle  
 (\frac{1}{\sqrt{6}}(2 0)
         +\frac{1}{\sqrt{2}}(1 0)+\frac{1}{\sqrt{3}}(0 0),1)
+ ( (2 +\!\!2),-1) 	 
  -(\frac{1}{\sqrt{2}}(2 +\!\!1)
         +\frac{1}{\sqrt{2}}(1 +\!\!1),0)
	 \bigg| V_{32}\nn \\
& \bigg|&( 
 (\frac{1}{\sqrt{6}}(2 0)
         +\frac{1}{\sqrt{2}}(1 0)+\frac{1}{\sqrt{3}}(0 0),1)
+ (2 +\!\!2),-1) 
  -(\frac{1}{\sqrt{2}}(2 +\!\!1)
         +\frac{1}{\sqrt{2}}(1 +\!\!1),0)\bigg\rangle.
\ea
where the notation followed in the last term for the states is 
$\langle (I^{\textrm{total}} I_z^{\textrm{total}},
I_z^k)|V_{ij}|\rangle$, where $I^{\textrm{total}}$ means the total
isospin of the $ij$ system and $k\neq i,j$ (the spectator $\rho$).

This leads, in terms of the $I=0$ and $I=2$ 
unitarized amplitudes ($t_{\rho\rho}^{(I=0)}$, $t_{\rho\rho}^{(I=2)}$),
to 
the following amplitude for the single scattering contribution:
\be
t_1=\frac{2}{9}\left(5t_{\rho\rho}^{(I=0)}+t_{\rho\rho}^{(I=2)}\right).
\ee
where we have added an extra 2 factor in order to match the unitary
normalization of ref.~\cite{Molina:2008jw} in $t_{\rho\rho}^{I}$.

It is worth noting that the argument of the function $T(s)$ is the total
invariant mass energy $s$, while the argument of $t_1$ is
$s'$, where $s'$ is the invariant mass of the
$\rho$ meson with momentum $k$ and the $\rho$ meson inside the $f_2$
resonance with momentum $p_1$ and is given by
\ba
s'=(k+p_1)^2=\frac{1}{2}\left(
s+3m_\rho^2-M_{f_2}^2\right)
\label{eq:sp12}
\ea 

For latter applications, let us write the general expression of $s'$ for
the interaction of a particle $A$ 
with a molecule
$B$  with $n$ equal building blocks $b$.
Then, $s'$ represents the invariant mass 
of the particle $A$ and a particle $b$ of the $B$ molecule
and is given by

\ba
s'=\frac{1}{n}\left(s-M_B^2-M_A^2\right)+M_A^2+m_b^2
\label{eq:spgeneral}
\ea
where $M_{A(B)}$ is the mass of the $A(B)$ system and $m_b$ is the
mass of every building block of the $B$ molecule.

Let us consider the wavefunctions of the incident and outgoing $\rho$
particles being plane waves normalized inside a box of volume ${\cal V}$
and let us call $\varphi_i$ the wavefunctions of the
$\rho$ mesons inside the $f_2$ resonance.
The $S$-matrix for the process of Fig.~\ref{fig:Faddeev}a is written as

\ba
S^{(1)}&=&\int d^4x
\frac{1}{\sqrt{2\omega_{p_1}}}e^{-ip_1^0x^0}\varphi_1(\vec x)
\frac{1}{\sqrt{2\omega_{p'_1}}}e^{i{p'}_1^0x^0}\varphi_1(\vec x)
\frac{1}{\sqrt{2\omega_k{\cal V}}}e^{-ikx}
\frac{1}{\sqrt{2\omega_k'{\cal V}}}e^{ik'x}
(-it_1)
\ea
which we can multiply by the identity
\be
\int d^3\vec x\,'\varphi_2(\vec x\,')\varphi_2(\vec x\,')=1.
\ee

The integration of the time component $x^0$ 
provides the energy conservation at the interaction point $x$:
\be
\int d x^0e^{-ip_1^0x^0}e^{i{p'}_1^0x^0}e^{-ik^0x^0}e^{ik'^0x^0}
=2\pi\,\delta(p_1^0+k^0-{p'}_1^0-k'^0)
\equiv 2\pi\,\delta(k^0+E_{f_2}-k'^0-E'_{f_2}),
\ee 
where in the last step we have assumed $p_2^0={p'_2}^0$, as corresponds to
having the second particle as spectator (impulse approximation).
We can take
\be
\varphi_1(x)\varphi_2(x')=\frac{1}{\sqrt{{\cal V}}}
e^{i\vec K_{f_2}\cdot \vec
R}\Psi_{f_2}(\vec r),
\ee
with $\Psi_{f_2}$ the wave function of the $f_2(1270)$ cluster and
\ba
\vec R=\frac{\vec x+\vec x'}{2} \nn\\
\vec r=\vec x-\vec x'.
\label{eq:changevarRr}
\ea
and then we get for the spatial integrals

\ba
\int d^3R\, 
e^{i\vec K_{f_2}\cdot \vec R} 
e^{-i\vec K'_{f_2}\cdot \vec R}
e^{i\vec k\cdot \vec R}
e^{-i\vec k'\cdot \vec R}
=(2\pi)^3\delta(\vec k+\vec K_{f_2}-\vec k'-\vec K'_{f_2})
\label{eq:Rintegral}
\ea
and
\ba
\int d^3r\, 
\Psi_{f_2}(\vec r)\Psi_{f_2}(\vec r)
e^{i\vec k\cdot \frac{\vec r}{2}}
e^{-i\vec k'\cdot \frac{\vec r}{2}}
= F_{f_2}\Big(\frac{\vec k -\vec k'}{2} \Big)\simeq
 F_{f_2}(0)=1
\ea
where $F_{f_2}$ is the $f_2(1270)$ form factor normalized to unity
neglecting the $\vec k$, $\vec k'$ momenta, which we take equal.

Hence the $S$-matrix for the single scattering term is given by

\ba
S^{(1)}&=&-it_1 \frac{1}{{\cal V}^2}
\frac{1}{\sqrt{2\omega_{p_1}}}
\frac{1}{\sqrt{2\omega_{p'_1}}}
\frac{1}{\sqrt{2\omega_k}}
\frac{1}{\sqrt{2\omega_k'}}
(2\pi)^4\,\delta(k+K_{f_2}-k'^0-K'_{f_2}).
\label{eq:Ssingle}
\ea

and recall we must sum $t_1+t_2 \to 2 t_1$.

\subsection{Double-scattering and resummation contribution}

We are going to evaluate the amplitude of the double-scattering
contribution (fig.~\ref{fig:Faddeev}b) in a similar way as in the case
of the kaon deuteron interaction in \cite{Kamalov:2000iy,Jido:2009jf}.

The $S$-matrix can be written as

\ba
S^{(2)}&=&\int d^4x\int d^4x'
\frac{1}{\sqrt{2\omega_{p_1}}}e^{-ip_1^0x^0}\varphi_1(\vec x)
\frac{1}{\sqrt{2\omega_{p'_1}}}e^{i{p'}_1^0x^0}\varphi_1(\vec x)
\frac{1}{\sqrt{2\omega_{p_2}}}e^{-ip_2^0x'^0}\varphi_2(\vec x\,')
\frac{1}{\sqrt{2\omega_{p'_2}}}e^{i{p'}_2^0x'^0}\varphi_2(\vec x\,')\nn\\
&&
\frac{1}{\sqrt{2\omega_k{\cal V}}}e^{-ikx}
\frac{1}{\sqrt{2\omega_k'{\cal V}}}e^{ik'x'}
i\int \frac{d^4q}{(2\pi)^4}
\frac{e^{iq(x-x')}}{q^2-m_\rho^2+i\epsilon}(-it_1)(-it_1)
\label{eq:Sdouble}
\ea

The integrations of the time components $x^0$ and ${x'}^0$
provide the energy conservation at the two interaction points $x$ and $x'$:
\ba
\int d x^0e^{-ip_1^0x^0}e^{i{p'}_1^0x^0}e^{-ik^0x^0}e^{iq^0x^0}
&=&2\pi\,\delta(p_1^0+k^0-{p'}_1^0-q^0)\nn\\
\int d {x'}^0e^{-ip_2^0{x'}^0}e^{i{p'}_2^0{x'}^0}e^{i{k'}^0{x'}^0}e^{-iq^0{x'}^0}
&=&2\pi\,\delta(p_2^0+q^0-{p'}_2^0-{k'}^0)
\ea 

We implement now the change of variables $(\vec x, \vec x')\to(\vec R, \vec r)$ 
of Eq.~(\ref{eq:changevarRr}).
The $R$ integral gives the same expression as 
in Eq.~(\ref{eq:Rintegral}), and the $\vec r$ integral gives rise to
\ba
&&\int d^3r\, 
\Psi_{f_2}(\vec r)\Psi_{f_2}(\vec r)
e^{i\vec k\cdot \frac{\vec r}{2}}
e^{i\vec k'\cdot \frac{\vec r}{2}}
e^{-i\vec q\cdot \vec r}\nn\\
&&=\int d^3r\, 
e^{-i( \vec q -\frac{\vec k +\vec k'}{2} )\cdot \vec r}
\,\Psi_{f_2}(\vec r)^2\equiv F_{f_2}\Big( \vec q -\frac{\vec k +\vec k'}{2} \Big)
\label{eq:Ff22}
\ea
where $F_{f_2}\left( \vec q -(\vec k +\vec k')/2 \right)$
is the $f_2(1270)$ form factor introduced above.

The final expression for the $S$-matrix for the double scattering process is
\ba
S^{(2)}&=&-i(2\pi)^4\delta(k+K_{f_2}-k'-K'_{f_2})\frac{1}{{\cal V}^2}
\frac{1}{\sqrt{2\omega_k}} 
\frac{1}{\sqrt{2\omega_k'}}
\frac{1}{\sqrt{2\omega_{p_1}}}
\frac{1}{\sqrt{2\omega_{p'_1}}}
\frac{1}{\sqrt{2\omega_{p_2}}} 
\frac{1}{\sqrt{2\omega_{p'_2}}}\nn\\
&&\times \int \frac{d^3q}{(2\pi)^3} 
F_{f_2}(q)
\frac{1}{{q^0}^2-\vec{q}\,^2-m_\rho^2+i\epsilon} t_1 t_1.
\label{eq:finalS2}
\ea
and we will take $q^0$ at the $f_2$ rest frame,
$q^0=(s-m_\rho^2-M_{f_2}^2)/(2M_{f_2})$,
where we have considered ${p_1}^0={p'_1}^0$ and ${p_2}^0={p'_2}^0$ which is true in
average. In Eq.~(\ref{eq:finalS2}) we have also taken into account
that  $(\vec k+\vec k')/2=0$ in average.

For the evaluation of the form factor of the $f_2$ resonance
we follow the approach of \cite{Gamermann:2009uq}. In this work it is shown that the use of a separable potential in momentum space of the type 
\be
V=v\theta(\Lambda -q)\theta(\Lambda -q')
\ee
  where $\Lambda$ is the cutoff used in the theory for the scattering of
 two particles and $q,q'$ are the modulus of the momenta, leads to the
 same on  shell prescription for the scattering matrix as is used in
 the  chiral unitary approach. The on shell prescription converts the
 coupled  integral equations for the scattering matrix into algebraic
 equations, and similarly, the wave functions can be easily obtained in
 terms of an integral. The wave function in momentum space is written as
\be \langle \vec{p}\,|\psi\rangle 
=v\,\frac{\Theta(\Lambda-p)}{E-\omega_\rho(\vec p_1)-\omega_\rho(\vec p_2)}\int_{k<\Lambda}
\dtres k \langle \vec{k}|\psi\rangle \label{wavep},
\ee 
where $\omega_\rho(\vec p)=\sqrt{\vec{p}\,^2+m_\rho^2}$, 
and in coordinate space as

\be \langle \vec{x}|\psi\rangle 
= \int \frac{\dtres p}{(2\pi)^{3/2}} e^{i\vec{p}.\vec{x}} \langle \vec{p}\,|\psi\rangle . \label{eq26}
\ee

The final expression for the form factor of Eq.~(\ref{eq:Ff22})
 is then given by

\ba
F_{f_2}(q)=\frac{1}{ {\cal N}}
\int_{\substack{p<\Lambda\\|\vec p-\vec q|<\Lambda}}
d^3p\,
\frac{1}{M_{f_2}-2\omega_\rho(\vec p)}\,
\frac{1}{M_{f_2}-2\omega_\rho(\vec p-\vec q)},
\label{eq:ff2cutoff}
\ea 
where the normalization factor ${\cal N}$ is
\ba
{\cal N}=
\int_{p<\Lambda}d^3p
\frac{1}{\left( M_{f_2}-2\omega_\rho(\vec p) \right)^2}.
\ea 

 \begin{figure}[!t]
\begin{center}
\includegraphics[width=0.6\textwidth]{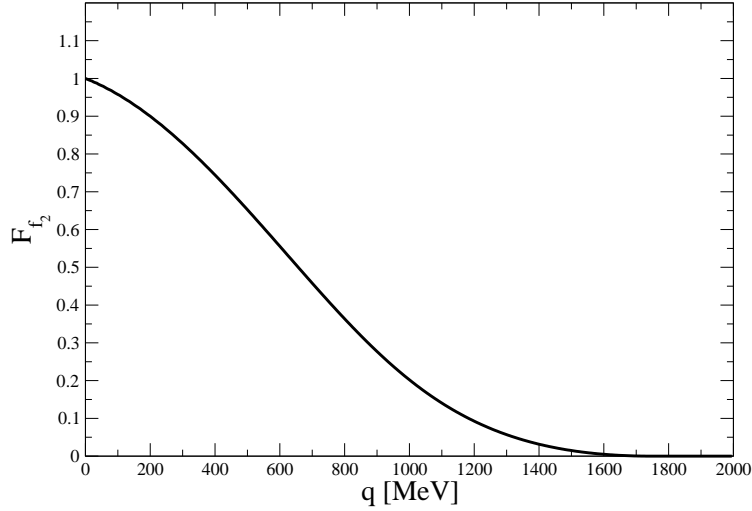}
\caption{Form factor of the $f_2(1270)$ resonance}
\label{fig:Ff2}
\end{center}
\end{figure}

In fig.~\ref{fig:Ff2} we show the form factor of the $f_2$ resonance.
 The condition 
$|\vec p-\vec q|<\Lambda$ implies that the form factor is exactly zero
for $q>2\Lambda$. Therefore the $d|\vec q|$ integration in
 Eq.~(\ref{eq:finalS2}) has an upper
limit of $2\Lambda$.

We must now face the issue of normalization in our formalism. We use
Mandl-Shaw \cite{mandl} normalization for the fields and hence the
$S$-matrix for $\rho f_2$ scattering is written as

\ba
S&=&-iT_{\rho f_2}(s)\frac{1}{ {\cal V}^2 }
\frac{1}{\sqrt{2\omega_k}}
\frac{1}{\sqrt{2\omega_{k'}}}
\frac{1}{\sqrt{2\omega_{f_2}}}
\frac{1}{\sqrt{2\omega_{{f_2}'}}}
(2\pi)^4\,\delta(k+K_{f_2}-k'^0-K'_{f_2})
\ea
but this should be compared with expressions Eq.~(\ref{eq:Ssingle}) for
the single scattering 
and Eq.~(\ref{eq:finalS2}) for double scattering. Summing the two
partitions $T_1$ and $T_2$ we find that
\be
T_{\rho f_2}=4 (t_1+ t_1 t_1 G_0),
\label{eq:4t1G0}
\ee
where we have made the assumption that in the $f_2$ rest frame, where we
evaluate the amplitude, $2\omega_\rho\simeq M_{f_2}$, and $G_0$ is given
by

\ba
G_0\equiv\frac{1}{M_{f_2}}\int\frac{d^3q}{(2\pi)^3} 
F_{f_2}(q)
\frac{1}{{q^0}^2-\vec{q}\,^2-m_\rho^2+i\epsilon}.
\label{eq:G0}
\ea

 \begin{figure}[!t]
\begin{center}
\includegraphics[width=0.6\textwidth]{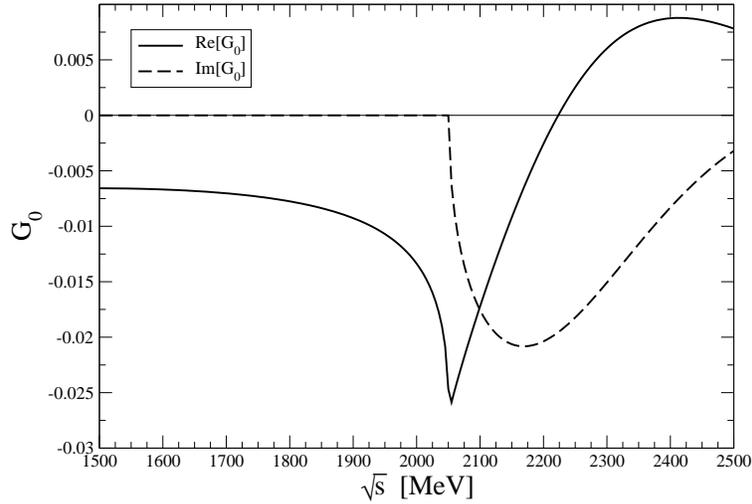}
\caption{Real and imaginary parts of the $G_0$ function, 
Eq.~(\ref{eq:G0})}
\label{fig:G0}
\end{center}
\end{figure}

In fig.~\ref{fig:G0} we show the real and imaginary parts of the $G_0$
function. Note that close to the threshold it has the typical
 shape of a two meson loop
function, $\rho f_2$ in this case,  but it is
smoothed towards zero at higher energies due to the form factor.

Equation~(\ref{eq:4t1G0}) represents the two first terms of the series
expansion of $4 t_1/(1-t_1G_0)$.
Actually, if we consider further number of scatterings in the expansion
of $T_{\rho f_2}$ of the FCA, (see diagrams $d$ in fig.~\ref{fig:Faddeev}), we get 
\ba
T_{\rho f_2}&=&4(t_1+t_1G_0 t_1+t_1 G_0 t_1G_0 t_1
+t_1 G_0 t_1G_0 t_1G_0 t_1
+...)=\frac{4t_1}{1-G_0 t_1}=\frac{4}{t_1^{-1}-G_0} \nn\\
&=&4 \left[t_1^{-1}(s')-\frac{1}{M_{f_2}}\int\frac{d^3q}{(2\pi)^3} 
F_{f_2}(q)
\frac{1}{{q^0}^2-\vec{q}\,^2-m_\rho^2+i\epsilon}\right]^{-1},
\ea 
where $s'$ is given in Eq.~(\ref{eq:sp12}).

\subsection{Larger number of $\rho$ mesons}

For the interaction of up to 6 $\rho $ mesons we can follow a similar
procedure as in the previous subsections but considering the
interaction of two different clusters. For the interaction of 4 $\rho$
meson we can calculate the interaction of 2$f_2(1270)$ resonances given
the strong tendency of two $\rho$ mesons to clusterize into an $f_2$.
Advancing some results that we will show later, this four $\rho$ state
gives rise to the $f_4$ resonance. Thus, analogously, for 5$\rho$ we
can consider the interaction of one $\rho$ meson with an $f_4$. And for
6$\rho$ we can consider the interaction of an $f_4$ with an $f_2$.

Therefore the amplitude for the interaction of a cluster $A$ 
with a cluster $B$ made of two equal components $b$ is given by

\ba
t(s;A,B)=4\left[t^{-1}(s'(s;A,b);A,b)-G_0(s;A,B)\right]^{-1}
\label{eq:tgeneral}
\ea
where
\ba
G_0(s;A,B)=\frac{1}{M_B}
\int\frac{d^3q}{(2\pi)^3} 
F(q;B)
\frac{1}{{q^0(s;A,B)}^2-\vec{q}\,^2-M_A^2+i\epsilon},
\ea

\ba
F(q,B)=\frac{1}{ {\cal N}}
\int_{\substack{p<\Lambda'\\|\vec p-\vec q|<\Lambda'}}
d^3p\,
\frac{1}{M_B-2\sqrt{\vec{p}\,^2+m_b^2}}\,
\frac{1}{M_B-2\sqrt{|\vec p-\vec q|^2+m_b^2}},
\ea 
\ba
{\cal N}=
\int_{p<\Lambda'}d^3p
\frac{1}{\left(M_B-2\sqrt{\vec{p}\,^2+m_b^2}\right)^2},
\ea 
\ba
q^0(s;A,B)=\frac{s-M_A^2-M_B^2}{2M_B}
\ea
and, from Eq.~(\ref{eq:spgeneral}),
\ba
s'(s;A,b)=\frac{1}{2}\left(s-M_B^2-M_A^2\right)+M_A^2+m_b^2.
\ea

Note that it is not necessary that the cutoff $\Lambda'$ be 
the same in all
the cases as the $\Lambda$ used for
the $f_2$ case.
The cutoff $\Lambda$ used in Eq.~(\ref{eq:Gcutoff}) for the $\rho\rho$
loop function, which is the same appearing in the momentum integral to
get the $f_2$ form factor in Eq.~(\ref{eq:ff2cutoff})
\cite{Gamermann:2009uq}, can be interpreted as
 the typical 
maximum momentum that each $\rho$ can reach inside the $f_2$ molecule.
For the $f_4$ case we can argue that the maximum value would be like in
the $f_2$ case but scaled by the typical momentum of the $f_2$
components inside the $f_4$ molecule. The typical three-momentum of the
 components
of the $f_4$, $\gamma_4$,
is of the order
of 
\ba
\gamma_4\sim\sqrt{\frac{B^2}{4}+M_{f_2}B}\qquad;\qquad 
B= M_{f_4}-2M_{f_2}
\ea
where $B$ is the binding energy of the $f_4$ and analogously for the
$f_2$:
\ba
\gamma_2\sim\sqrt{\frac{B^2}{4}+m_{\rho}B}\qquad;\qquad 
B= M_{f_2}-2m_{\rho}
\ea

This gives for the cutoff of the $f_4$
\ba
\Lambda'\bigg|_{f_4}\sim \Lambda
\sqrt{\left|\frac{\gamma_4}{\gamma_2}\right|}\simeq 1500\mev.
\label{eq:cutoff1500}
\ea
While this is just a very rough estimation, this gives us an idea of the
order of $\Lambda'$. In any case this only affects the
evaluation of the $5\rho$ and $6\rho$ system.
In the numerical evaluation we will consider the range 
$\Lambda'\sim 875-1500\mev$ to have an idea of the uncertainties
from this source. But, advancing some results, the dependence of the
mass of the systems with this cutoff is small.

\subsection{Arbitrary number of $\rho$ mesons in single scattering
approximation}

For and arbitrary number of $\rho$ mesons, it is possible to obtain a
simple analytic expression for the mass of the multi-$\rho$ system
if only the single scattering mechanism is
considered, which is the first order mechanism. Of course this 
is just a toy approximation since, as we
will see in the results section, the multiple scattering is important,
but it serves to
make some interesting qualitative arguments.

In the single scattering approximation, Eq.~(\ref{eq:tgeneral}) takes
the form
\ba
t(s;A,B)\simeq4\, t(s'(s;A,b);A,b)
\label{eq:tgeneralsing}
\ea

That means that, for instance, the amplitude $t_{\rho f_2}$ is just
proportional to the $t_{\rho\rho}$ amplitude but evaluated at an energy
$s'$ shifted with respect to $s$  due to the fact that one of the
$\rho$'s involved in the $\rho\rho$ scattering 
 is bound into an $f_2$ system. In general, the interaction amplitude
of   a number $n_\rho$ of $\rho$ mesons is proportional to the
$\rho\rho$ amplitude with an energy obtained considering that one of
the $\rho$ mesons is bound into an $(n_\rho-1)$ molecule. Therefore one
can obtain recursively the amplitude for the $n_\rho$ system.
Because of that, the shape of $|t(s;A,B)|^2$ is the same as that of 
$|t_{\rho\rho}(\tilde s)|^2$ but at a shifted energy. The $\tilde s$
value at which $|t_{\rho\rho}(\tilde s)|^2$ has the maximum is precisely
$M_{f_2}^2$.
 The value of $s$ appearing in $t(s;A,B)$
 of Eq.~(\ref{eq:tgeneralsing})
  is the value that we can assign to the
mass of the $n_\rho$ system, $M(n_\rho)$. Therefore, applying
recursively the above condition one can obtain a general expression for
$M(n_\rho)$ in the single scattering approximation:

\ba
M(n_\rho)^2=\frac{1}{2}n_\rho\left(n_\rho-1\right)M_{f_2}^2
-n_\rho\left(n_\rho-2\right)m_\rho^2.
\label{eq:Mrhon}
\ea

We can also define a binding energy per $\rho$ as
\ba
E(n_\rho)=\frac{n_\rho m_\rho-M(n_\rho)}{n_\rho}
\label{eq:Erhon}
\ea
which will be used for later discussions.

\section{Results}
\label{sec:results}

\begin{figure*}
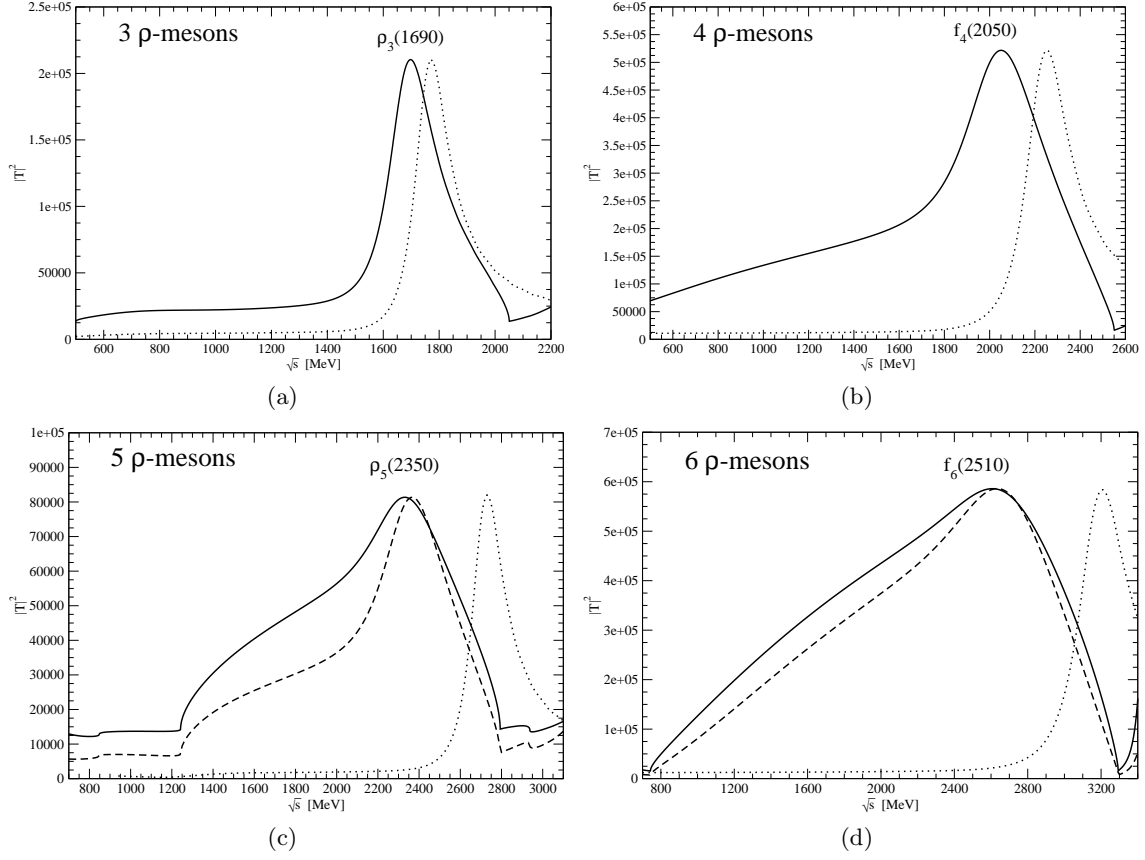

     \centering
      \subfigure[]{
          \label{fig:T2_rho3}
          \includegraphics[width=.45\linewidth]{figure5a.eps}}
      \subfigure[]{
          \label{fig:T2_f4}
          \includegraphics[width=.45\linewidth]{figure5b.eps}}
       \subfigure[]{
          \label{fig:T2_rho5}
          \includegraphics[width=.45\linewidth]{figure5c.eps}}
      \subfigure[]{
          \label{fig:T2_f6}
          \includegraphics[width=.45\linewidth]{figure5d.eps}}
   \caption{Modulus squared of the unitarized multi-$\rho$ amplitudes.
   Solid line: full model $\Lambda'\big|_{f_4}=1500$\mev;
   dashed line: full model $\Lambda'\big|_{f_4}=875$\mev;
       dotted line: only single-scattering
   contribution. (The dashed and dotted lines have been normalized to
   the peak of the solid line for the sake of comparison 
   of the position 
   of the maxima)}
     \label{fig:T2s}
\end{figure*}

In fig.~\ref{fig:T2s} we show the modulus squared of the amplitudes for
different number of $\rho$ mesons considering only the single scattering
mechanisms (dotted line) and the full model (solid and dashed lines).
The difference between the solid and dashed lines is the value of 
$\Lambda'\big|_{f_4}$ of Eq.~(\ref{eq:cutoff1500}) 
needed in the evaluation of the
$5\rho$ and $6\rho$ meson systems
(1500~MeV 
in the
solid line, 875~MeV in the dashed one).
The
dotted and dashed curves have been normalized to the peaks of the
corresponding full result for the sake of comparison of the position of
the maximum.
The difference between the dashed and solid lines can be
considered as an
estimate of the error but the variation in the position of the maximum
is small.

We clearly see that the amplitudes show  
pronounced bumps which we associate to the resonances 
labeled in the figures.
The position of the 
the maxima can be associated to the masses of the corresponding
resonances.

In table~\ref{tab:masses} the values of the masses of our generated
multi-$\rho$ systems are shown in comparison with the experimental
values at the PDG \cite{Amsler:2008zzb}. The two values for the
$\rho_5$ and $f_6$ masses in the full model column correspond to the
different values in $\Lambda'\big|_{f_4}$ as explained above.
In the last column the binding energy per $\rho$ meson, 
$E(n_\rho)=(n_\rho m_\rho-M(n_\rho))/n_\rho$ is also shown.

\begin{table}[h]
\begin{center}
\begin{tabular}{|c|c|c|c|c|c|} 
\hline
 $n_\rho$ &  & mass, PDG \cite{Amsler:2008zzb} 
& mass, only single scatt. & mass,
 full model & $E(n_\rho)$  \\ \hline
2 & $f_2(1270)$    & $1275\pm  1$ & 1275   & 1285	 & 133 \\ \hline
3 & $\rho_3(1690)$ & $1689\pm  2$ & 1753   & 1698	 & 209 \\ \hline
4 & $f_4(2050)$    & $2018\pm 11$ & 2224   & 2051	 & 263 \\ \hline
5 & $\rho_5(2350)$ & $2330\pm 35$ & 2690   & 2330-2366   & 302-309 \\ \hline
6 & $f_6(2510)$    & $2465\pm 50$ & 3155   & 2607-2633   & 337-341 \\ \hline
 \end{tabular}
\end{center}
\caption{Results for the masses of the dynamically generated states.}
\label{tab:masses}
\end{table}
In fig.~\ref{fig:Mvsn} we show graphically the results for the masses 
of 
table~\ref{tab:masses}.

 \begin{figure}[!t]
\begin{center}
\includegraphics[width=0.7\textwidth]{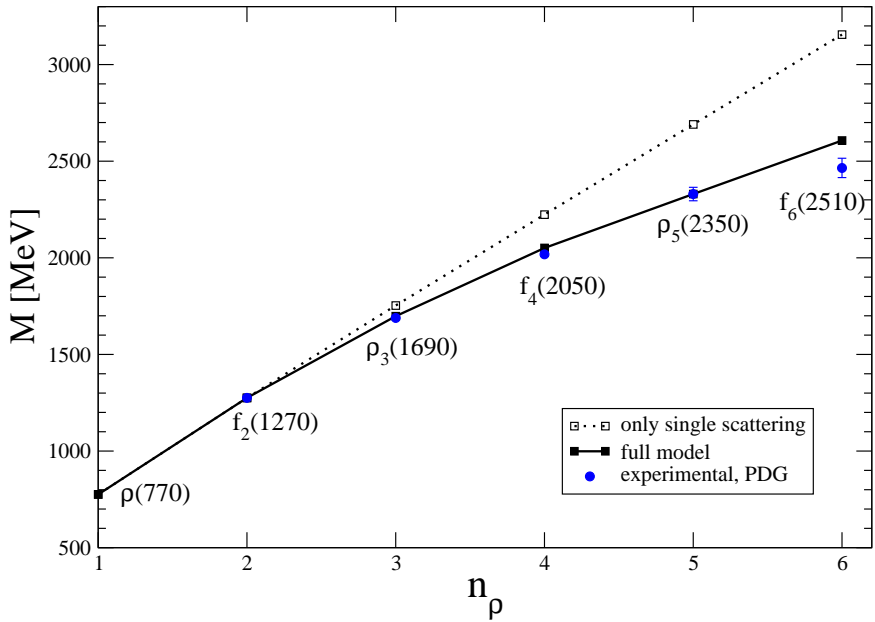}
\caption{Masses of the dynamically generated states as a function of the
number of constituent $\rho(770)$ mesons, $n_\rho$. Only single
scattering contribution (dotted line); full model (solid line);
experimental values from the PDG\cite{Amsler:2008zzb}, (circles).
}
\label{fig:Mvsn}
\end{center}
\end{figure}

We can see from the results that 
the single scattering mechanism produce
qualitatively the resonances but the positions of the masses do not
agree with the experimental values by differences ranging from about
$60\mev$ for the $\rho_3$ to $700\mev$ for the $f_6$. The situation is
drastically improved when the multiple scattering is considered.  In
this case, the agreement with the experimental values of the masses is
remarkable. Quantitatively, the full model is essentially 
compatible with the
experimental values within errors except for the $f_6$ where the
discrepancy is about $150\mev$, which is still quite remarkable, given
the high mass and width of the resonance.
The typical discrepancy with the experimental masses is of
the order of 1\%, (5\% for the $f_6$).

It is worth stressing the simplicity of our approach 
and the absence of
parameters fitted in the model. To be more precise, only the value
$\Lambda=875\mev$ of the  cutoff of the $\rho\rho$ loop function
was chosen in ref.~\cite{Molina:2008jw} to agree with the experimental
$f_2$ pole position. No further adjustments have been done in the
present work. 

In principle, the widths of the bumps can be associated to the 
the widths of the resonances if they were Breit-Wigner like shapes,
 which is clearly not the case. This means that the amplitudes contain much
non-resonant background which our model generates implicitly through the
non-linear dynamics involved in the unitarization procedure. That means
that the extraction of the widths of the resonances from our amplitudes 
is just very qualitative: 200, 350, 900 and $1500\mev$ for $\rho_3$, $f_4$,
$\rho_5$ and $f_6$ respectively.
 The order of magnitude agree with the
experimental value of the PDG \cite{Amsler:2008zzb},
 $161\pm 10$, $237\pm 18$, $400\pm 100$
and $255\pm 40$ respectively, except for the two heaviest states. 
However,
it is worth noting that, by looking at fig.~\ref{fig:T2s},
in these heaviest states much of the strength of the amplitude off the
peak could be interpreted as a background, as would be the case in an
experimental analysis of a distribution like the one obtained in 
fig.~\ref{fig:T2s}, in
 which case the actual width of the resonance would be significantly
reduced.

Let us address again the problem of an arbitrary large number of
$\rho$ mesons. A natural question looking at fig.~\ref{fig:Mvsn}
is if the curve of the masses saturates for a large enough
 number of $\rho$
mesons. That would imply that it would be energetically free to add an
extra $\rho$ meson to the system.
For the single scattering case we can
analyze the problem with the help of Eq.~(\ref{eq:Mrhon}). The
saturation would occur for 
\be
n_\rho\Big|_{\textrm{sat}}=\frac{m_\rho^2-M_{f_2}^2/4}
{m_\rho^2-M_{f_2}^2/2}
\label{eq:nsat}
\ee
which never happens for the actual $\rho$ and $f_2$ masses, since
$n_\rho\Big|_{\textrm{sat}}$ of Eq.~(\ref{eq:nsat}) gives a negative
value.
However it is 
worth noting in fig.~\ref{fig:Mvsn}
that the single scattering is just a bound limit and that the multiple
scatterings tend to decrease importantly the mass of an  $n_\rho$
system.
If such decrease is enough to eventually reach the saturation condition
 cannot be
answered with certainty within the present model 
since we do not go beyond $n_\rho=6$. However, it is worth noting
 the large value of 
the binding energy per $\rho$, see last column
of table~\ref{tab:masses}.  Already for $n_\rho=6$, it is almost
 half
the value of the $\rho$ meson mass. That means that the creation of
the 6 $\rho$
meson system gives back half the mass of all the particles involved 
which is quite a lot of energy. 

The binding energy per $\rho$ evaluated using
the single scattering approximation, 
Eq.~(\ref{eq:Erhon}), tends
asymptotically to 
\ba
\lim_{n_\rho\to\infty}E(n_\rho)
=m_\rho-\sqrt{\frac{M_{f_2}^2}{2}-m_\rho^2}\simeq 315 \textrm{ MeV}.
\ea 
However this value is already reached at $n_\rho=6$ if the multiple
scattering mechanisms are considered.

If the $\rho\rho$ interaction were a little bit stronger, such that
$M_{f_2}\sim \sqrt{2}m_\rho=1096\mev$, then the saturation would be
reached already considering only the single scattering. And, in order
to get saturation for  $n_\rho=6$ with only single scattering, 
the mass of the $f_2$ resonance 
should be just slightly smaller, $\sim\sqrt{20/11}m_\rho=1056\mev$.
Of course this is just a qualitative reasoning
 since the width of the system
would eventually increase with the number of $\rho$ mesons, making the
system fade away rapidly. 
The former discussion is obviously rough and speculative of what might
happen for large  $n_\rho$ systems. What remains as quantitative results
from the present study is the fact that
the $f_2(1270)$, $\rho_3(1690)$,   $f_4(2050)$,
 $\rho_5(2350)$ and   $f_6(2510)$ can be essentially considered as 
multi-$\rho$ molecules with increasing number of $\rho$ mesons.

\section{Conclusions}
\label{sec:concl}

In the present work we claim for the first time that
the $\rho_3(1690)$,
$f_4(2050)$, $\rho_5(2350)$ and $f_6(2510)$ resonances 
can be interpreted as
multi-$\rho$ states of 3, 4, 5 and 6 $\rho$ mesons respectively, with
their spins aligned. 
The main
idea stems from the fact that in
ref.~\cite{Molina:2008jw} it was found that the interaction
of two $\rho(770)$ mesons in isospin $I=0$ and spin $S=2$ is very
strong, to the point to bind the two $\rho$ mesons forming
the $f_2(1270)$ resonance. This elementary $\rho\rho$ interaction
is obtained
implementing unitarity, using the techniques of the chiral unitary
approach, with a potential  obtained from a hidden gauge symmetry
Lagrangian for the interaction of two vector
mesons. For the multi-$\rho$ systems we evaluate the scattering
amplitudes for the interactions of two clusters made up of
$\rho$-mesons. To this purpose we use 
the fixed center approximation to the Faddeev equations which considers
the multiple scattering steps in addition to the single process where each
$\rho$ meson interacts with all the rest of $\rho$ mesons within the
cluster. 

The position of the maximum in the modulus squared of the amplitudes can be associated
with the masses of the corresponding resonances. It is worth noting
that the model has no free parameters once a cutoff is chosen 
in ref.~\cite{Molina:2008jw}
to obtain the experimental mass of the $f_2(1270)$ resonance.

The values of the masses that we obtain are in very good
 agreement with the
experimental values of the masses of the resonances considered in the
present work, the $\rho_3(1690)$,
$f_4(2050)$, $\rho_5(2350)$ and $f_6(2510)$. This is a remarkable fact given the simplicity of the
underlying idea.

The states obtained have an increasing binding energy per particle which
induces one to speculate on the possibility that for a given number of
$\rho$ mesons it would cost no energy to produce a new $\rho$ meson
inside the meson condensate state. However,
simultaneusly we observe that the width of the new systems also
increases with the number of $\rho$ mesons,
 to the point that for $n_\rho=6$
the width is already very large. It might as well be that one has
reached an experimental threshold and that new multi $\rho$ states, that
in principle could be created, have such a large width that they escape
present detection techniques.  In any case, the claims made here that the
already observed states up to  $J=6$ correspond to multi $\rho$ states
is a novel idea worth consideration. New studies with different
formalisms and different points of view would be most welcome, as well as
possible experimental tests which could help unveil the real nature of
these states.

\section*{Acknowledgments}
We wish to thank Raquel Molina for providing us the $\rho \rho$ amplitudes.
This work is partly supported by DGICYT contracts  FIS2006-03438,
FPA2007-62777 and 
the EU Integrated Infrastructure Initiative Hadron Physics
Project  under Grant Agreement n.227431.

\end{document}